\documentclass[letterpaper]{article} 
\usepackage{aaai25}  
\usepackage{times}  
\usepackage{helvet}  
\usepackage{courier}  
\usepackage[hyphens]{url}  
\usepackage{graphicx} 
\usepackage{natbib}  
\usepackage{caption} 
\usepackage{algorithm}
\usepackage{algorithmic}

\usepackage{graphicx}
\usepackage{framed}
\usepackage{paralist}
\usepackage{booktabs}
\usepackage{multirow}
\usepackage{array, makecell}
\usepackage{pifont}
\usepackage{xcolor}
\newcommand{\answerYes}[1]{\textcolor{blue}{#1}} 
\newcommand{\answerNo}[1]{\textcolor{teal}{#1}} 
\newcommand{\answerNA}[1]{\textcolor{gray}{#1}}

\newlength{\DepthReference}
\newlength{\HeightReference}
\newcommand{\MyColorBox}[2][red]%
{%
	\settowidth{\Width}{#2}%
	\colorbox{#1}%
	{%
		\raisebox{-\DepthReference}%
		{%
			\parbox[b][\HeightReference+\DepthReference][c]{\Width}{\centering#2}%
		}%
	}%
}
\usepackage{newfloat}
\usepackage{listings}
\DeclareCaptionStyle{ruled}{labelfont=normalfont,labelsep=colon,strut=off} 
\floatstyle{ruled}
\newfloat{listing}{tb}{lst}{}
\floatname{listing}{Listing}
\title{Measuring and Forecasting Conversation Incivility: \\
	the Role of Antisocial and Prosocial Behaviors}
\author{
	Xinchen Yu\textsuperscript{\rm 1},
	Hayden Arnold\textsuperscript{\rm 2},
	Benjamin Su\textsuperscript{\rm 3},
	Eduardo Blanco\textsuperscript{\rm 1}
}
\affiliations{
	\textsuperscript{\rm 1}University of Arizona\\	
	\textsuperscript{\rm 2}Korea Advanced Institute of Science and Technology\\
	\textsuperscript{\rm 3}University of Southern California\\
	\{xinchenyu, eduardoblanco\}@arizona.edu, harnold@kaist.ac.kr, subenjam@usc.edu
}

\usepackage{bibentry}

\begin{document}
	
	\maketitle
	
	\begin{abstract}
		This paper focuses on the task of measuring and forecasting incivility in conversations following replies to hate speech.
		Identifying replies that steer conversations away from hatred and elicit civil follow-up conversations sheds light into effective strategies to engage with hate speech and proactively avoid further escalation.
		We propose new metrics that take into account various dimensions of antisocial and prosocial behaviors to measure the conversation incivility following replies to hate speech.
		Our best metric aligns with human perceptions better than prior work.
		Additionally, we present analyses on a) the language of antisocial and prosocial posts, b) the relationship between antisocial or prosocial posts and user interactions, and c) the language of replies to hate speech that elicit follow-up conversations with different incivility levels.
		We show that forecasting the incivility level of conversations following a reply to hate speech is a challenging task.
		We also present qualitative analyses to identify the most common errors made by our best model.

	\end{abstract}
	
	%
	\section{Introduction}
	\label{sec:introduction}
	Online discussion platforms enable new forms of interaction and have democratized public discourse on an immense scale.
	Hate speech, however, challenges their benefits. 
	It harms individuals who suffer from personal attacks~\cite{olteanu2018effect}, distracts people from the goals of discussions~\cite{arazy2013stay}, and even influences offline hate crimes~\cite{10.1145/3292522.3326045}.
	Engaging with hate speech, for example by rebutting hateful content or discouraging hateful speakers, has emerged as a promising approach to address this problem.
	Compared with reactive moderation in which ``bad actors" and ``objective content" are identified and removed~\cite{chang2022thread}, 
	properly engaging with hate speech may divert the discourse away from hatred
	and avoid escalating tense situations.
	Further, engagement does not restrict free speech~\cite{schieb2016governing}.
	
	\begin{figure}[t!]
		\centering
		\includegraphics[width=1\linewidth]{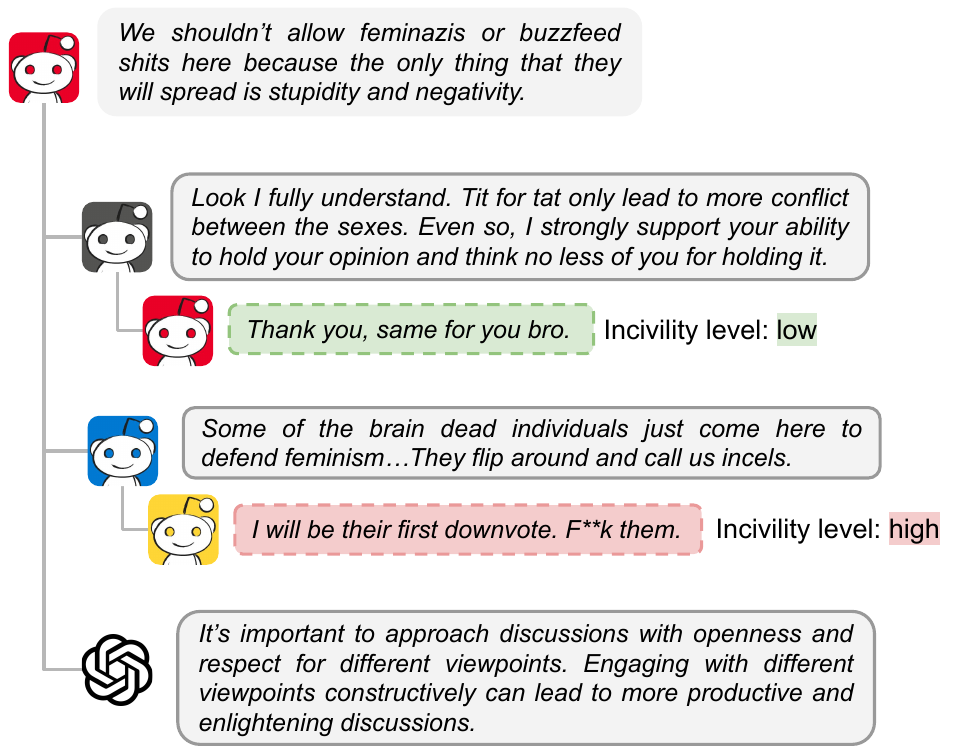}
		\caption{Hateful Reddit post (top), three direct replies, and the follow-up conversations.
			The first reply steers the follow-up conversation towards civil behaviors.
			The second reply elicits additional incivility.
			The last reply is generated by ChatGPT.
			It does not address the hateful post directly or elicit a follow-up conversation.
		} 
		\label{fig:illustration}
	\end{figure}
	
	Prior work has focused on modeling replies to hate speech, including corpora construction~\cite{mathew2019thou, chung-etal-2019-conan}, fine-grained categorization~\cite{mathew2019thou, yu-etal-2023-fine}, and  generation~\cite{zhu-bhat-2021-generate,gupta-etal-2023-counterspeeches,chung-bright-2024-effectiveness}.
	Still, the effectiveness of replies to hate speech is understudied. 
	When engaging with hate speech, conversational outcomes vary depending on the strategies used.
	Consider the Reddit conversation in Figure~\ref{fig:illustration}.
	There are three direct replies to the hateful post:
	two from actual Reddit users and one from a generative model, ChatGPT~\cite{chatgpt2024}.
	The first reply disagrees with the hateful post showing empathy and support---two prosocial behaviors.
	It successfully steers the author of the hateful post towards civil behavior in the follow-up conversation.
	The second reply uses denigrating language---an antisocial behavior---and results in an additional uncivil post. 
	The third reply is polite but not as rich as genuine replies written by Reddit users. 
	The effectiveness of such generic replies remains unclear.
	
	In this paper, we forecast the incivility of conversations following replies to hate speech in user-generated content.
	Doing so opens the door to identify content that is likely---and unlikely---to result in escalation of hatred
	before this undesirable outcome becomes a reality.
	The forecasting proposed here could be used to direct
	moderation efforts to conversations where human intervention is most needed to reduce hatred.
	At the same time, the forecasting would also 
	minimize moderation when users' replies to hateful posts organically elicit civil conversations.
	Beyond real-world applications,
	our work provides data-driven observations into the language most effective at discouraging incivility
	as well as strategies to navigate difficult conversations.
	
	We take a broad view of conversation incivility by considering
	a combination of antisocial, prosocial, and neutral behaviors. 
	Antisocial behaviors have been defined as acts intended to harm or disadvantage another individual~\cite{sage2006goal},
	whereas prosocial behaviors are acts intended to help or benefit another person~\cite{eisenberg1998handbook}.
	Different from traditional approaches that model antisocial or prosocial behaviors separately~\cite{zhang-etal-2018-conversations,bao2021conversations,lambert2022conversational}, 
	we make the first attempt to jointly model both.
	Inspired by the fact that prosocial behaviors are the opposite of antisocial behaviors~\cite{bar1976prosocial} but not a dichotomy,
	we examine whether their presence in a conversation is effective at estimating conversation incivility.
	Notably, instead of modeling a single dimension, for example only considering norm violations for antisocial behaviors, we work with diverse dimensions.
	As concepts of antisocial and prosocial behaviors are complex~\cite{10.1145/3232676,bao2021conversations}, we argue that including different dimensions could capture the nuances and help us understand which aspects contribute most to the perceptions of conversation incivility.
	
	
	This study draws its conclusions from a large Reddit dataset.
	We start by studying the conversations following replies to hate speech from three complementary perspectives:
	(a) presence of prosocial and antisocial behaviors discussed in the literature,
	(b) linguistic features of prosocial and antisocial behaviors,
	and
	(c) user interactions.
	Then, we analyze the role of antisocial and prosocial behaviors in measuring conversation incivility.
	Human validation demonstrates that combining several dimensions of both kinds of behaviors results in a more robust metric than prior work.
	Finally, we experiment with classifiers to forecast the incivility of the conversation following a reply to hateful content.
	The experimental results show that the task is challenging, and
	we close with an error analysis.
	
	In summary, our main contributions are:
	\begin{compactitem}
		\item New metrics that take into account several dimensions of both antisocial and prosocial behaviors in measuring conversation incivility following replies to hate speech;
		\item Comparing different types of replies to hate speech with respect to language usage, including a) the difference between antisocial replies and prosocial replies and b) the difference between replies eliciting conversations with different incivility levels: high, medium and low;
		\item Analyzing antisocial and prosocial behaviors in user interactions with regard to two scenarios: re-engagement and multi-turn conversations;
		\item Building models to predict conversation incivility level and presenting a qualitative error analysis.
	\end{compactitem}

	\section{Related work}
	\label{sec:relatedwork}
	
	\noindent
	\textbf{Antisocial Behavior} 
	Prior work has studied a wide range of antisocial behaviors in online platforms such as Reddit~\cite{vidgen-etal-2021-introducing}, Instagram~\cite{liu2018forecasting}, and Wikipedia~\cite{zhang-etal-2018-conversations}.  
	These antisocial behaviors include hate speech~\cite{rottger-etal-2021-hatecheck}, abusive language~\cite{vidgen-etal-2021-introducing}, offensive language~\cite{zampieri-etal-2019-predicting}, toxicity~\cite{pavlopoulos-etal-2020-toxicity}, and norm violations~\cite{lambert2022conversational}.
	Most of these previous studies focus on reactive moderation, that is, detecting antisocial behaviors after they have occurred.  
	Our work, similar to another line of prior research~\cite{zhang-etal-2018-conversations,lambert2022conversational}, 
	aims at forecasting whether (future) conversations will be uncivil.
	We focus on forecasting incivility of conversations following replies to hateful posts.

	\noindent
	\textbf{Prosocial Behavior} 
	Existing work on prosocial behaviors has explored politeness~\cite{danescu-niculescu-mizil-etal-2013-computational}, empathy~\cite{buechel-etal-2018-modeling, liu2024measuring}, sympathy and encouragement~\cite{sosea-caragea-2022-ensynet}, positiveness~\cite{ziems-etal-2022-inducing}, donation~\cite{dong-etal-2022-text} and norms~\cite{ziems-etal-2023-normbank}.
	Specifically, \citet{bao2021conversations} and \citet{lambert2022conversational} have proposed various prosocial metrics to quantify and predict prosocial outcomes in follow-up conversations.
	Instead of using hand-crafted lexicon features, we leverage deep neural networks to identify prosocial behaviors.
	To the best of our knowledge,
	we are the first to examine whether prosocial and antisocial behaviors
	even out the incivility people perceive in a conversation.
	
	\noindent
	\textbf{Forecasting Conversational Incivility} 
	There have been several efforts on predicting whether an event will lead to derailing future conversations.
	This includes predicting future removal of a post~\cite{cheng2017anyone}, personal attacks~\cite{zhang-etal-2018-conversations}, and incivility intensity~\cite{liu2018forecasting,dahiya2021would,Yu_Blanco_Hong_2024}.
	Given a conversation, prior work uses different metrics to estimate incivility. 
	\citet{liu2018forecasting} considers the number of uncivil posts.
	\citet{dahiya2021would} averages the scores of comments from a toxicity classifier.
	\citet{Yu_Blanco_Hong_2024} takes into account the number of both uncivil and civil comments as well as the unique users involved in the conversation.
	Our paper extends these works by combining several dimensions of both antisocial and prosocial comments for the first time.
	Doing so results in a metric that more closely resembles human judgments.

	\section{An Analysis of Reddit Conversations Following Replies to Hateful Posts}
	\label{sec:corpus}
	We first describe 
	(a) the procedure to collect a large collection of relevant Reddit conversations (hateful post, reply, and follow-up conversation)
	and 
	(b) several dimensions of antisocial and prosocial behaviors from the literature.
	Then, to better understand conversations following replies to hateful posts,
	we conduct analyses on such conversations containing prosocial and antisocial behaviors.
	Specifically, we investigate 
	(a) linguistic characteristics to reveal differences in language use
	and
	(b) user interactions (e.g., re-engagement, multi-turn conversations).
	
	\paragraph{Dataset}
	Our starting point is the Reddit dataset collected by \citet{Yu_Blanco_Hong_2024}.
	It consists of 1,382,596 posts from 39 subreddits and binary labels indicating whether each post is hateful.
	Reddit, known for its large size of user populations and diverse topics~\cite{baumgartner2020pushshift},
	is an ideal source for studying online conversations.
	We first select all the hateful posts and create (\emph{hateful post}, \emph{reply}, \emph{follow-up conversation}) triples by pairing (a) each hateful post with each of its direct replies and (b) each direct reply to the hateful post with the follow-up conversation.
	This strategy results in 25,225 hateful posts, 41,727 replies to the hateful posts, and a total of 124,172 posts in the follow-up conversations.
	After removing replies that have been moderated, we obtain 38,041 triples.
	Each triple consists of one hateful post, one reply, and the follow-up conversation.
	Follow-up conversations can be empty and the maximum length is 975 posts.

	\paragraph{Identifying Antisocial and Prosocial Comments} 
	
	\begin{figure}[t!]
		\centering
		\includegraphics[width=0.9\linewidth]{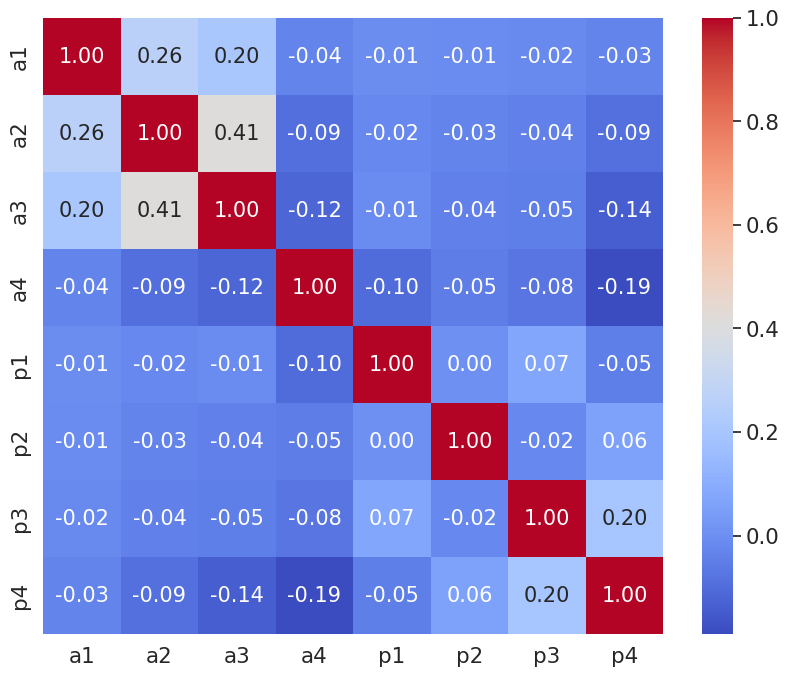}
		\caption{Spearman's rank correlation coefficients between antisocial and prosocial behaviors.
			$a_i$ and $p_i$ indicate antisocial and prosocial behaviors in the order described in the paper.
			All coefficients are low ($< 0.3$)
			except the one between $a_2$~(explicit hate speech) and $a_3$~(abusive language).
		} 
		\label{fig:classifiers}
	\end{figure}
	
	We work with four antisocial behaviors and four prosocial behaviors drawn from the literature.
	Our selection criteria includes not only theoretical background,
	but also whether corpora exist with which to train classifiers for large-scale automatic annotation.
	For antisocial behaviors, we work with the following:
	(a)~offensive language: offensive or not offensive~\cite{davidson2017automated},
	(b)~explicit hate speech: hateful or not hateful~\cite{qian-etal-2019-benchmark},
	(c)~abusive language: abusive or not abusive~\cite{vidgen-etal-2021-introducing},
	and 
	(d)~norm violations (whether the content of a post is ``\emph{deleted}" or ``\emph{removed}"): yes or no.
	For prosocial behaviors, we work with the following:
	(a)~empathy: expressing (including both weak and strong communications) or not expressing~\cite{sharma-etal-2020-computational},
	(b)~norms: expected or not expected~\cite{ziems-etal-2023-normbank},
	(c)~positiveness: positive or not positive~\cite{ziems-etal-2022-inducing},
	and 
	(d)~politeness: polite or not polite~\cite{danescu-niculescu-mizil-etal-2013-computational}.
	
	We fine-tune a RoBERTa transformer~\cite{liu2019robertarobustlyoptimizedbert} for each dimension of antisocial and prosocial behavior in order to build a classifier.
	See Appendix \ref{a:classifierperformance} for the experimental results.
	Then, we use the resulting classifiers to indicate
	whether each post in the conversation following a reply to hate speech
	exhibits each antisocial and prosocial behavior.
	
	A perhaps obvious question is whether the four antisocial and prosocial behaviors account for the same phenomena
	and thus differ only in name.
	We refer the reader to the original papers for details, but the answer is that that is not the case.
	Figure~\ref{fig:classifiers} shows the correlation coefficients between all antisocial and prosocial behaviors considered as observed in the follow-up conversations to replies to hateful posts.
	All the correlations are low ($< 0.3$) except the one between \emph{explicit hate speech} and \emph{abusive language}.
	In other words, each antisocial and prosocial behavior captures a distinct characteristic in a Reddit post.

	\subsection{The Language of Antisocial and Prosocial Posts}
	\label{subsec:linguistic}
	
	\begin{table}[t]
		\centering
		\small
		\begin{tabular}{lc}
			\toprule
			& p-value \\
			\midrule
			Textual factors \\
			~~~~~~~~First pronoun  & \colorbox{red!10}{$\uparrow\uparrow\uparrow$}  \\
			~~~~~~~~Second pronoun  & \colorbox{blue!10}{$\downarrow\downarrow\downarrow$} \\
			~~~~~~~~Total tokens    &\colorbox{red!10}{$\uparrow\uparrow\uparrow$}  \\
			~~~~~~~~Negation cues   &\colorbox{red!10}{$\uparrow\uparrow\uparrow$}  \\
			~~~~~~~~Question mark   &\colorbox{red!10}{$\uparrow\uparrow\uparrow$} \\
			\midrule
			Sentiment factors \\
			~~~~~~~~Disgust words  &\colorbox{red!10}{$\uparrow\uparrow\uparrow$}  \\
			~~~~~~~~Sadness words  &\colorbox{red!10}{$\uparrow\uparrow\uparrow$}  \\
			~~~~~~~~Negative words  &\colorbox{red!10}{$\uparrow\uparrow\uparrow$} \\			
			~~~~~~~~Positive words   &\colorbox{blue!10}{$\downarrow\downarrow\downarrow$}  \\
			~~~~~~~~Happiness words   &\colorbox{blue!10}{$\downarrow\downarrow\downarrow$}  \\
			~~~~~~~~Gratitude words   &\colorbox{blue!10}{$\downarrow\downarrow\downarrow$} \\
			~~~~~~~~Hostile words  &\colorbox{red!10}{$\uparrow\uparrow\uparrow$}  \\
			~~~~~~~~Angry words   &\colorbox{red!10}{$\uparrow\uparrow\uparrow$} \\
			\bottomrule
		\end{tabular}
		\caption{Linguistic analysis comparing antisocial and prosocial posts. 
			Three arrows indicates p-value~\textless~0.001 (unpaired t-test).
			Arrow direction indicates whether higher values correlate with antisocial posts (up) or  prosocial posts (down).}
		\label{t:linguistic-anti-pro}
	\end{table}
	
	While it is well-known that prosocial behaviors are the opposite of antisocial behaviors~\cite{bar1976prosocial},
	the actual differences between these behaviors remains unknown.
	To find out the language differences,
	we compare linguistic features of antisocial and prosocial posts in the conversations following replies to hate.
	We consider a post antisocial if it is identified as such by any antisocial classifier, 
	and similarly, 
	we consider a post prosocial if it is identified as such by any prosocial classifier.
	We exclude posts identified as norm violations in this analysis, as their content is always ``\emph{deleted}."
	This results in 25,312 antisocial post and 25,260 prosocial posts. 
	We consider both textual features (e.g., first and second pronouns, negation cues) and sentiment features~\cite{crossley2017sentiment}.
	Negation cues checks for presence in the list by \citet{fancellu-etal-2016-neural}.
	We also conduct the Bonferroni correction as multiple hypothesis tests are performed.
	
	Table~\ref{t:linguistic-anti-pro} shows the results that passed the Bonferroni correction.
	Regarding textual features, people use significantly~($p < 0.001$) more first pronouns, tokens,  negation cues and question marks when attacking and harming other individuals (antisocial behavior). 
	There are more second pronouns~(e.g., \emph{you} and \emph{your}) in prosocial posts, indicating caring behaviors~\cite{hoffman1996empathy}.
	Regarding sentiment features, there are significantly~($p < 0.001$) more negative, disgust, hostile and angry words in antisocial comments.
	On the other hand, we find more happiness, gratitude, and positiveness in prosocial comments.

	\subsection{Do Antisocial Behaviors and Prosocial Behaviors Influence User Interactions?}
	\label{subsec:interaction}
	To understand whether antisocial and prosocial behaviors
	affect how people interact (e.g., re-engagement, multi-turn conversations),
	we analyze the structure of the follow-up conversations after replies to hateful posts.
	We work with all the posts in the follow-up conversations and answer the following questions.
	
	\paragraph{Does re-engagement differ after receiving antisocial or prosocial posts?}
	Yes, it does. 
	We focus on users who receive both antisocial and prosocial posts (4,537)
	and compare their re-engagement in the subsequent conversations.
	We analyze re-engagement from two perspectives:
	how often they re-enter after each antisocial (or prosocial) post
	(a) anywhere in the subsequent conversation
	and
	(b) immediately after the antisocial (or prosocial) post (i.e., direct replies).
	Results show that when receiving antisocial posts,
	the percentage of re-engagement---including both anywhere and immediately after the antisocial post---is significantly higher than when receiving prosocial comments (paired sample t-test, $p < 0.001$).
	
	\paragraph{Are there more antisocial posts than prosocial posts in multi-turn conversations?}
	No, there are not.
	We consider multi-turn conversations those in which there are at least two turns between each pair of users in a conversation.
	We conduct paired sample t-tests to see which behaviors (antisocial or prosocial) are more frequent in multi-turn conversations.
	Results show that prosocial posts are significantly more frequent than antisocial posts ($p<0.001$),
	despite the percentage of antisocial posts (20.38\%) is slightly higher than prosocial posts (20.34\%) in all the follow-up conversations.
	In other words, the amount of antisocial posts is roughly the same in the follow-up conversations,
	but there are significantly more prosocial posts in multi-turn conversations---the longer the conversation, the better the tone.

	\paragraph{Do people who engage in a multi-turn conversation display the same behavior? }
	No, they do not.
	We examine whether two people in a multi-turn conversation behave similarly.
	That is, whether they both display either antisocial or prosocial behavior or a mix.
	For each pair of users in a multi-turn conversation,
	we calculate the percentage of antisocial and prosocial posts out of all posts directed at each other.
	Then, we calculate the difference in percentages of antisocial (\emph{anti\_diff}) and prosocial (\emph{pro\_diff}) per pair of users.
	Finally, we calculate the difference between \emph{anti\_diff} and \emph{pro\_diff} per pair.
	If both users have similar behaviors, the final difference will be small.
	On the other hand, if a user displays more antisocial (or prosocial) behaviors than the other, the absolute final difference would be large.
	Results show that there is significant difference (one-sample t-test, $p<0.001$)
	in two users' behaviors in multi-turn conversations.
	In other words, it is common to have conversations in which two users display the opposite behaviors---antisocial and prosocial behaviors do not elicit the same behavior from a second speaker in multi-turn conversations.
	
	\section{Measuring Conversation Incivility}
	\label{sec:metric}
	We aim to define an automatic metric to assess conversation incivility so that low-cost, large-scale annotations become a reality.
	Unlike previous work which considers either \emph{one} antisocial or \emph{one} prosocial behavior,
	our proposal considers
	(a)~both antisocial and prosocial behaviors as well as neutral behaviors
	and
	(b)~four types of antisocial and prosocial behaviors.
	By experimenting with several weighting mechanisms, we replicate all previous metrics.
	Careful evaluation allows us to conclude that our metric outperforms previous proposals.
	In fact, when our automated metric is considered an ``annotator'' to identify which of two follow-up conversations is more uncivil,
	it obtains inter-annotator agreement
	(compared to a ``real'' (human) annotator) above the threshold to be considered reliable.
	

	\begin{figure}[t]
		\centering
		\includegraphics[width=1\linewidth]{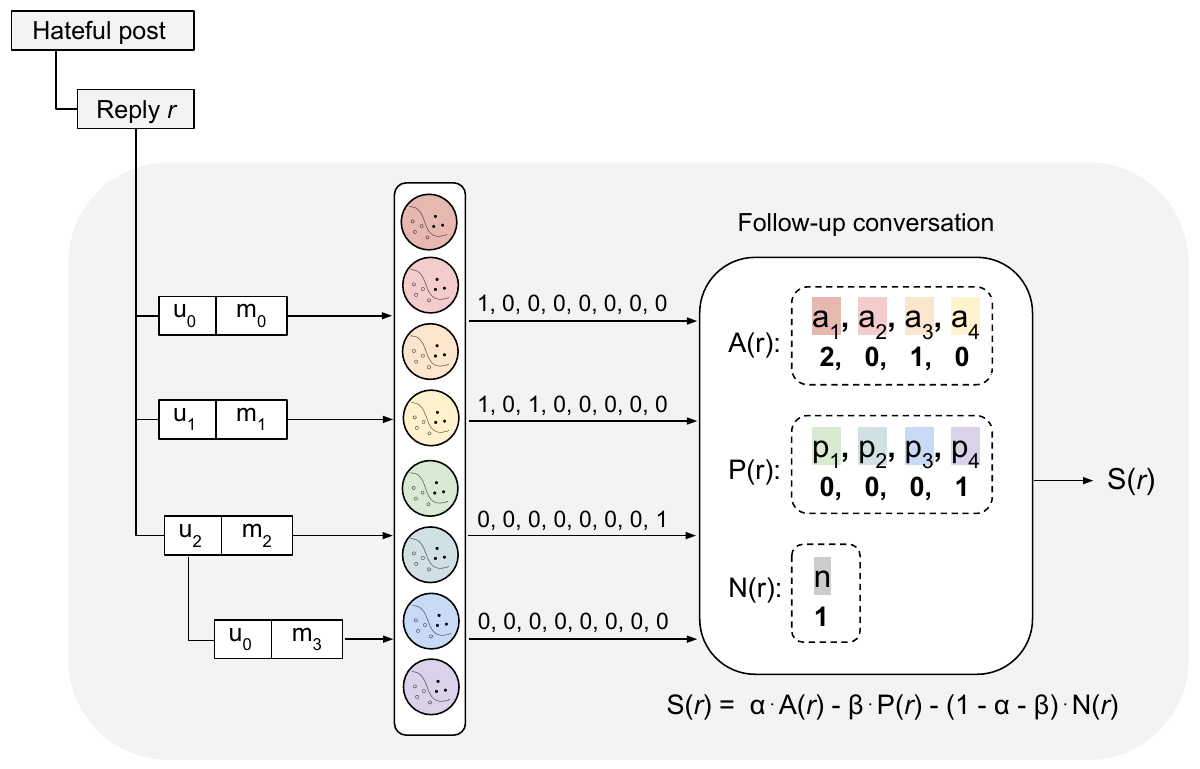}
		\caption{
			Illustration of our metric to estimate incivility of the conversation following a reply $r$ to a hateful post ($S(r)$).
			We calculate four antisocial and four prosocial behaviors for each post.
			The metric consists of components to account for 
			antisocial ($A(r)$),
			prosocial ($P(r)$),
			and neutral behaviors ($N(r)$);
			and considers not only the amount of each behavior but also whether different authors generate posts with the same behavior (not shown).
		}
		\label{fig:metric}
	\end{figure}
	
	\begin{figure*}[t!]
		\centering
		\includegraphics[width=0.7\linewidth]{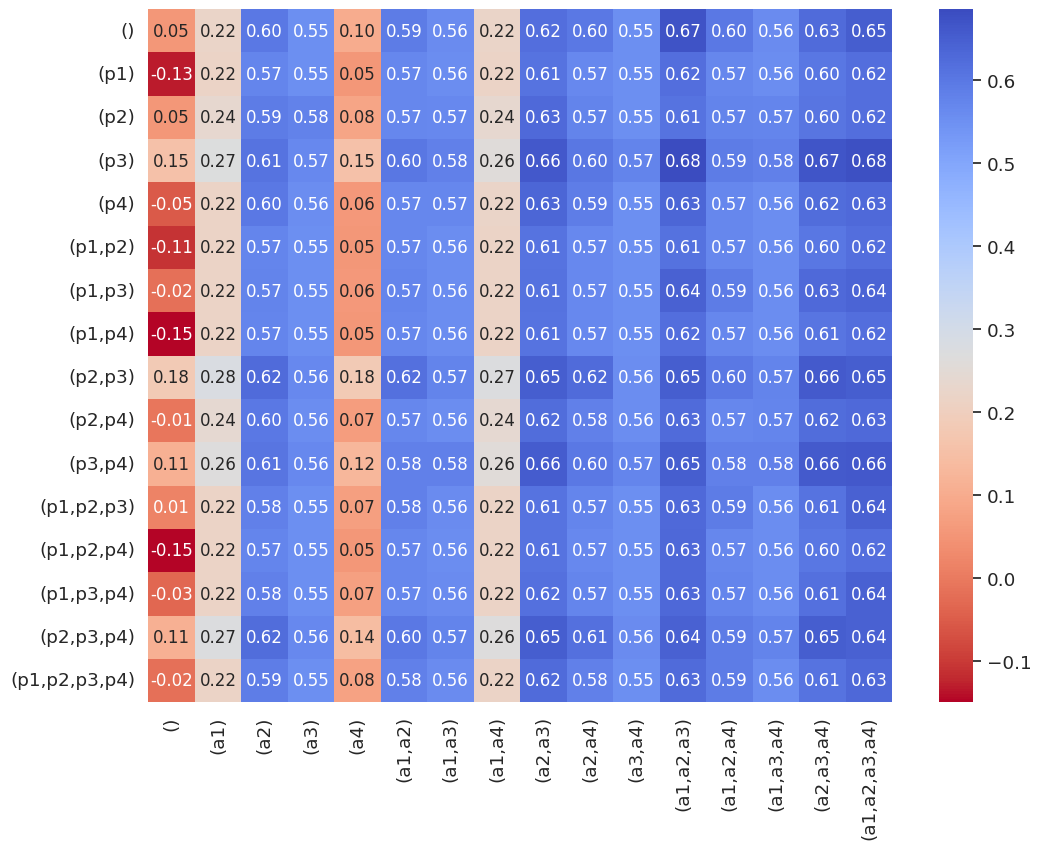}
		\caption{Cohen's $\kappa$ coefficients between human annotations (after adjudication) and
			using our metric to determine which of two conversations is more uncivil.
			Cells indicate the highest $\kappa$ obtained with the corresponding combination of antisocial and prosocial behaviors after trying all combinations of $\alpha$ and $\beta$.
			Prosocial behaviors by themselves underperform,
			and combining several antisocial behaviors outperforms individual antisocial behaviors.
			The optimal choice is $(a_1, a_2, a_3)$ and $(p_3)$.
		} 
		\label{fig:kappas}
	\end{figure*}

	\paragraph{A Metric to Measure Conversation Incivility}
	Our metric includes three main components accounting for antisocial, prosocial, and neutral behavior.
	Further, we include user re-entry behaviors~\cite{backstrom2013characterizing} to identify whether the same author engages in multiple posts with the same behavior.
	Intuitively, we give less weight to the same author displaying the same behavior in multiple posts compared to the same amount of posts by several authors.

	Given a reply $r$ to a hateful post,
	the conversation incivility score of the follow-up conversation ($S(r)$)
	consists of three components: 
	antisocial component $A(r)$,
	prosocial component $P(r)$,
	and
	neutral component $N(r)$.
	The antisocial component consists of up to the four dimensions introduced earlier:
	$a_1$ (offensive language),
	$a_2$ (explicit hate speech),
	$a_3$ (abusive language),
	and $a_4$ (norm violations).
	Similarly, the prosocial component also consists of up to  four dimensions:
	$p_1$ (empathy),
	$p_2$ (norms),
	$p_3$ (positiveness),
	and $p_4$ (politeness).
	The neutral component includes posts that display neither antisocial nor prosocial behaviors.
	We assign equal weights to the antisocial and prosocial dimensions.
	For example, if we only consider \emph{offensive language} ($a_1$)
	and
	\emph{abusive language} ($a_3$) as indicators of incivility, a weight $0.5$ is assigned to each of the two dimensions.

	In addition to the amount of posts displaying the antisocial and prosocial behaviors, user re-entry is an important factor in measuring incivility~\cite{backstrom2013characterizing}.
	As pointed out by~\citet{Yu_Blanco_Hong_2024} within the hate and counterhate domain,
	the same amount of posts displaying a behavior should be given more weight if they come from several users as opposed to a single, prolific user.
	We adapt their insight to account for several prosocial and antisocial dimensions.
	Specifically, for each user $u$ ($u = 0,1,...,k$), we calculate the number of antisocial and prosocial posts $c_{a_i, u}$ and $c_{p_i, u}$ per dimension $i$ ($i = 1, 2, 3, 4$) as well as the count of neutral comments $c_{n, u}$. 
	A function~$f$ (e.g., square root) is applied on $c_{a_i, u}$, $c_{p_i, u}$, and $c_{n, u}$
	to differentiate instances in which the same user makes several posts with the same behavior (e.g., with $f = \sqrt{}{}$, 10 antisocial posts by 10 users are considered as uncivil as 100 posts by the same person).
	The choice of $f$ is not crucial, as the metric is better suited to be used in relative terms (i.e., for comparison purposes) rather than absolute terms.
	Summing up over all users in the conversation, we have: $A(r) = \sum_{u=0}^{k} f({c_{a_i, u}})$,
	$P(r) = \sum_{u=0}^{k} f({c_{p_i, u}})$,
	and
	$N(r) = \sum_{u=0}^{k} f({c_{n, u}})$.
	
	\begin{table*}[t]
		\centering
		\small
		\begin{tabular}{llccccc}
			\toprule
			& & \multicolumn{3}{c}{High vs Low} &  \multirow{1}{*}{\makecell{High vs Medium}}  &   \multirow{1}{*}{\makecell{Low vs Medium}}  \\
			\cline{3-4} 	\cline{6-6} 	\cline{7-7}
			\addlinespace[2pt]
			& & hateful post & reply && reply & reply   \\
			\midrule
			\multirow{6}{*}{\makecell{Textual factors}} & 1st person pronouns & $\uparrow\uparrow$ &  \colorbox{red!10}{$\uparrow\uparrow\uparrow$} &  & \colorbox{red!10}{$\uparrow\uparrow\uparrow$} &   \colorbox{red!10}{$\uparrow\uparrow\uparrow$} \\
			& 2nd person pronouns & \colorbox{red!10}{$\uparrow\uparrow\uparrow$} &  \colorbox{red!10}{$\uparrow\uparrow\uparrow$} & &  \colorbox{red!10}{$\uparrow\uparrow\uparrow$} &   \colorbox{red!10}{$\uparrow\uparrow\uparrow$} \\
			& Tokens & $\uparrow$ & \colorbox{red!10}{$\uparrow\uparrow\uparrow$} & & \colorbox{red!10}{$\uparrow\uparrow\uparrow$}  & \colorbox{red!10}{$\uparrow\uparrow\uparrow$}  \\
			& Negation cues & \colorbox{red!10}{$\uparrow\uparrow\uparrow$} & \colorbox{red!10}{$\uparrow\uparrow\uparrow$}  & & \colorbox{red!10}{$\uparrow\uparrow\uparrow$}  & \colorbox{red!10}{$\uparrow\uparrow\uparrow$}  \\
			& Quotations & $\uparrow\uparrow$ & \colorbox{red!10}{$\uparrow\uparrow\uparrow$} & & \colorbox{red!10}{$\uparrow\uparrow\uparrow$}  & \\
			& Question marks & \colorbox{red!10}{$\uparrow\uparrow\uparrow$} & \colorbox{red!10}{$\uparrow\uparrow\uparrow$}  & & \colorbox{red!10}{$\uparrow\uparrow\uparrow$}  & \colorbox{red!10}{$\uparrow\uparrow\uparrow$}  \\
			\midrule
			\multirow{6}{*}{\makecell{Sentiment factors}} & Disgust words & $\uparrow\uparrow\uparrow$ & \colorbox{red!10}{$\uparrow\uparrow\uparrow$} & & \colorbox{red!10}{$\uparrow\uparrow\uparrow$} & \\
			& Sadness words  & $\uparrow\uparrow\uparrow$ & \colorbox{red!10}{$\uparrow\uparrow\uparrow$} & & \colorbox{red!10}{$\uparrow\uparrow\uparrow$} & \\
			& Positive words & \colorbox{blue!10}{$\downarrow\downarrow\downarrow$} &   \colorbox{blue!10}{$\downarrow\downarrow\downarrow$} & &  \colorbox{blue!10}{$\downarrow\downarrow\downarrow$}  & \colorbox{red!10}{$\uparrow\uparrow\uparrow$}  \\
			& Negative words & \colorbox{red!10}{$\uparrow\uparrow\uparrow$}  & \colorbox{red!10}{$\uparrow\uparrow\uparrow$}  & &\colorbox{red!10}{$\uparrow\uparrow\uparrow$} & \\
			& Anger words & & \colorbox{red!10}{$\uparrow\uparrow\uparrow$} & & \colorbox{red!10}{$\uparrow\uparrow\uparrow$} & \\
			& Hostile words & $\uparrow\uparrow$ & \colorbox{red!10}{$\uparrow\uparrow\uparrow$} & & \colorbox{red!10}{$\uparrow\uparrow\uparrow$} & \\
			\bottomrule
		\end{tabular}
		\caption{Linguistic analysis comparing
			(a) hateful posts that result in high and low incivility in the follow-up conversations to their replies (Column 3)
			and 
			(b) replies to hateful posts that result in high, medium and low incivility in the follow-up conversation (Columns 4--6).
			Number of arrows indicates the p-value (t-test; one: $p<0.05$, two: $p<0.01$, and three: $p<0.001$).
			Arrow direction indicates whether higher values correlate with the first group (up or the second group (down) in each pairwise comparison. 
			Tests that pass the Bonferroni correction have background color.}
		\label{t:conversational-outcomes}
	\end{table*}

	The more antisocial posts, the more uncivil a conversation is.
	Similarly, the more prosocial and neutral posts,
	the more civil a conversation is.
	We thus define $S(r)$ as follows:
	$$S(r) = \alpha \cdot A(r) - \beta \cdot P(r) - (1-\alpha-\beta) \cdot N(r)$$
	Parameters $\alpha$ and $\beta$ are weights for the antisocial and prosocial components (and the neutral component).
	Figure~\ref{fig:metric} illustrates the procedure to calculate $S(r)$.

	The specific antisocial and prosocial dimensions as well as $\alpha$ and $\beta$ are parameters that must be tuned.
	We choose $\alpha, \beta \in [0, 1]$  with $\alpha + \beta \leq 1$. 
	We experiment with $0.05$ increments and compare against human judgments as explained below.
	Our data-driven metric is the first to consider both antisocial and prosocial behaviors, and, importantly, several dimensions of each.
	Note that our approach subsumes all prior metrics to measure conversation incivility.
	By trying all combinations of antisocial and prosocial behaviors,
	we subsume previous efforts considering either one antisocial or prosocial behavior~\cite{liu2018forecasting}.
	We also subsume efforts considering
	either antisocial ($\alpha=1$ and $\beta=0$) or prosocial ($\alpha=0$ and $\beta=1$) behaviors~\cite{bao2021conversations},
	antisocial and non-antisocial behaviors ($\alpha \in (0, 1)$ and $\beta = (1-\alpha)/2$),
	and
	prosocial and non-prosocial behaviors ($ \beta \in (0,1)$ and $\alpha = (1-\beta)/2$)~\cite{lambert2022conversational}.
	

	\paragraph{Tuning the Metric}
	Our metric definition must be tuned to determine
	the optimal antisocial and prosocial behaviors as well as $\alpha$ and $\beta$.
	While one could make decisions inspired by the literature, 
	we argue that a data-driven approach is more sound.
	To this end, we conduct human annotations so that we can 
	identify the optimal choices by comparing how closely the metric estimates human assessments.
	
	First, we randomly select two
	(hateful post, reply, follow-up conversations) triples
	accounting for a variety of lengths in the follow-up conversations.
	Specifically, we divide follow-up conversations into short~($\le 5$ posts),
	medium~($> 5$ posts and $\le 10$ posts)
	and long~($>10$ posts).
	Then, we generate 40 pairs of triples where the follow-up conversation belong to the following lengths: 
	\emph{short} and \emph{short}, 
	\emph{short} and \emph{medium},  
	\emph{short} and \emph{long}, 
	\emph{medium} and \emph{medium},  
	\emph{medium} and \emph{long}, and
	\emph{long} and \emph{long}.
	Finally, we shuffle the triples in the pair.
	The first step ensures that we
	have a representative sample of follow-up conversations as far as length.
	Thus, the tuning process is designed to result in a robust metric regardless of the length of the follow-up conversation.
	
	Second, we employ two native annotators to manually annotate which follow-up conversation in the triple is more uncivil in a pair.
	The interface displays the full triple (hateful post, reply, and follow-up conversation), similar to the illustration shown in Figure \ref{fig:illustration}.
	The Cohen's $\kappa$ is 0.84 (90\% accuracy) among the 240 pairs, which is considered nearly perfect~\cite{artstein2008inter}.
	
	
	Third, we try all combinations of antisocial and prosocial behaviors as well as $\alpha$ and $\beta$ to find the optimal choices.
	Armed with the manually annotated benchmark,
	this task is trivial.
	Note that any instantiation of our metric ($S(r)$)
	can be used to tell which of two conversations is more uncivil by comparing their incivility scores.
	Figure~\ref{fig:kappas} provides the Cohen's $\kappa$ coefficients comparing the results obtained with our metric and the human annotators.
	For each combination of antisocial and prosocial dimensions, the figure presents the highest $\kappa$ after trying all combinations of $\alpha$ and $\beta$.
	We make the following observations from the results:
	
	\begin{table*}[t]
		
		\small
		\centering	
		\begin{tabular}{l ccc ccc ccc ccc}
			\toprule
			\multicolumn{1}{c}{} & \multicolumn{3}{c}{High} & \multicolumn{3}{c}{Medium} & \multicolumn{3}{c}{Low} & \multicolumn{3}{c}{Weighted Average} \\
			\cmidrule(lr){2-4} \cmidrule(lr){5-7} \cmidrule(lr){8-10} \cmidrule(lr){11-13} 
			& P & R & F1 & P & R & F1 & P & R & F1 & P & R & F1 \\
			\midrule
			\addlinespace[2pt]
			Majority Baseline & 0.00 & 0.00 & 0.00 & 0.52 & 1.00 & 0.68 & 0.00 & 0.00 & 0.00 &  0.27 & 0.52 & 0.35 \\ 
			\midrule
			RoBERTa, training with \\ 		
			~~~reply & 0.38	& 0.52	& 0.44	& 0.60	& 0.70	& 0.64	& 0.33	& 0.10	& 0.15	& 0.48	& 0.51	& 0.47\\		
			~~~~~~+ pretraining† & 0.45	& 0.38	& 0.41	& 0.58	& 0.83	& \textbf{0.68}	& 0.34	& 0.09	& 0.14	& 0.49	& 0.54	& 0.49\\
			~~~~~~+ blending† & 0.43	& 0.49	& 0.46	& 0.60	& 0.73	& 0.66	& 0.35	& 0.15	& 0.21	& 0.50	& 0.53	& 0.50\\ 
			\addlinespace
			
			~~~hate + reply &0 .39 & 0.45 & 0.42 & 0.60 & 0.62 & 0.61 & 0.32 & 0.26 & \textbf{0.29} & 0.48 & 0.49 & 0.49\\
			~~~~~~+ pretraining†‡ & 0.42 & 0.53 &\textbf{ 0.47} & 0.61 & 0.72 & 0.66 & 0.38 & 0.15 & 0.22 & 0.51 & 0.53 & \textbf{0.51}\\
			~~~~~~+ blending†‡ & 0.44 & 0.51 & \textbf{0.47} & 0.62 & 0.72 & 0.66 & 0.34 & 0.17 & 0.23 & 0.51 & 0.53 & \textbf{0.51}\\ 
			\midrule
			GPT-4o, prompting with \\ 
			~~~reply, Zero-Shot & 0.30 & 0.59 & 0.40 & 0.52 & 0.36 & 0.43 & 0.24 & 0.18 & 0.21 & 0.40 & 0.37 & 0.37\\
			~~~reply, Few-Shot & 0.36 & 0.40 & 0.38 & 0.52 & 0.51 & 0.51 & 0.26 & 0.25 & 0.25 & 0.42 & 0.42 & \textbf{0.42}\\
			~~~hate + reply, Zero-Shot  & 0.29 & 0.56 & 0.38 & 0.53 & 0.45 & 0.48 & 0.25 & 0.11 & 0.16 & 0.40 & 0.39 & 0.38\\
			~~~hate + reply, Few-Shot & 0.33 & 0.42 & 0.37 & 0.50 & 0.47 & 0.48 & 0.26 & 0.24 & 0.25 & 0.40 & 0.40 & 0.40\\
			\bottomrule
			
		\end{tabular}
		\caption{Results obtained with several models. 
			We indicate statistical significance (McNemar’s test \cite{mcnemar1947note} over the weighted average) as follows: 
			† indicates statistically significant ($p<0.05$) results with respect to the \emph{reply} model,
			and ‡ with respect to the \emph{hate + reply} model.
			Training with the \emph{hate + reply}
			coupled with pretraining with stance or both pretraining or blending counterspeech yields the best results (F1: 0.51).}
		\label{t:model-results}
	\end{table*}
	
	\begin{compactitem}
		\item The best metric
		takes into account both antisocial ($a_1,a_2,a_3$) and prosocial behaviors ($p_3$) with $\alpha=0.75$ and $\beta=0.15$.
		It obtains a Cohen's $\kappa$ of $0.68$, which is  considered \emph{substantial}
		agreement~\cite{artstein2008inter}.
		While lower than the ``true'' inter-annotator agreement ($\kappa = 0.84$), our metric is considered a reliable annotator ($\kappa = 0.68$; coefficients between 0.6 and 0.8 are considered \emph{substantial} agreement).
		
		\item Antisocial behaviors are much more informative than prosocial behaviors.
		Indeed, considering only antisocial behaviors (top row) yields $\kappa = 0.67$.
		
		\item Considering several antisocial behaviors is beneficial. Indeed, single antisocial behavior yields $\kappa=0.22-0.60$, while considering more than one, yields $\kappa=0.67$.
		
		\item While prosocial behaviors obtain poor coefficients by themselves (first column), they are modestly beneficial when combined with antisocial behaviors.
		
	\end{compactitem}
	
	\paragraph{Categorizing Conversation Incivility}
	Having identified the best metric, using it to annotate the incivility score of all the follow-up conversations in our corpus is straightforward.
	We use the incivility scores to group all the replies to hateful posts based on the incivility quartiles in the follow-up conversations to the replies~(top 25\%, middle 50\%, and bottom 25\%).
	The exact score ranges are as follows:
	\begin{compactitem}
		\item \emph{Low} incivility: $S(r) \in (-16.38, -0.10]$;
		\item \emph{Medium} incivility: $S(r) \in (-0.10, 0]$; and
		\item \emph{High} incivility: $S(r) \in (0, 7.81]$.
	\end{compactitem}

	\section{Corpus Analysis}
	\label{sec:corpusanalysis}
	
	The final corpus consists of 38,041 (\emph{hateful post}, reply, follow-up conversation) triples with the incivility scores of the follow-up conversations.
	23.48\% of the replies are followed by conversations with high incivility, 
	25.43\% are followed by conversations with low incivility,
	and the rest (51.09\%) are followed by conversations with medium incivility.
	We note that 92.10\% of the replies with medium conversation incivility scores have no follow-up conversations. 
	This motivates us to explore the replies to hateful posts
	from two perspectives.
	First, we investigate the language in the replies that is likely to elicit follow-up conversations regardless of incivility.
	Second,
	we explore the differences in language between follow-up conversations with high and low incivility.
	We therefore compare the linguistic differences between replies to hate speech that a) have and do not have follow-up conversations (high vs. medium, low vs. medium), and
	b) have different incivility levels of follow-up conversations when they are not empty (high vs. low).
	We run unpaired t-tests and report results in Table~\ref{t:conversational-outcomes}.
	We observe several interesting findings:
	\begin{compactitem}
		\item Regarding textual features, replies that use more personal pronouns (both 1st and 2nd person), tokens, negation cues and question marks are likely to attract future conversations to follow, and at the same time, more incivility in these follow-up conversations.
		\item Regarding sentiment factors, there are significantly more negativeness, disgust, sadness, anger, and hostility in the replies followed by conversations with high incivility as well as less positiveness. 
		Besides the replies, the hateful posts that contain more negativeness and less positiveness elicit more incivility in the follow-up conversations.
	\end{compactitem}
	
	\section{Experiments}
	\label{sec:experiments}
	We experiment with models to determine the incivility level (i.e., high, medium, or low) of the conversation following a reply to the hateful post.
	We randomly split the 38,041 instances as follows:
	70\% for training,
	15\% for validation and 
	15\% for testing.
	To investigate whether taking into account the hateful posts is beneficial, we consider two textual inputs:
	a) the reply to the hateful post (reply), and
	b) the hateful post and the reply (hate + reply).
	We experiment with supervised approaches and prompting LLMs.
	
	\subsection{Supervised Approaches}
	We start with the off-the-shelf RoBERTa transformer~\cite{liu2019robertarobustlyoptimizedbert} released by Hugging Face~\cite{wolf-etal-2020-transformers}.
	We also experiment with FLAN-T5~\cite{chung2024scaling}. 
	As the performance is very close to RoBERTa, 
	we detail the results with FLAN-T5 in Appendix~\ref{a:experimentdetails}.
	We explore two strategies to improve performance of models.
	
	\begin{table*}
		\small
		\centering
		\begin{tabular}{p{2.5cm}rp{8.5cm}rr}
			\toprule
			Error Type & \% & Example & Prediction & Gold  \\ \midrule
			
			Rhetorical  question & 26 & Hate: \emph{I just enjoy calling out retards on the internet, it's not a crime.} & & \\
			&    & Reply: \emph{Maybe go look in a mirror?} & low & high   \\
			\midrule
			Irony or sarcasm & 18 & Hate: \emph{That's stupid and you're stupid.} & & \\
			&   &  Reply: \emph{I'd make a rebuttal but I hate getting into fights with children.} & low & high \\
			\midrule
			Swear words in & 13 & Hate: \emph{Saying all this during Ramadan too? You're a shitty Muslim.}  &  &\\
			the reply &  & Reply: \emph{From people like you who find it normal to abuse us. You are a hypocrite, pathetic, just another sheep. You are a shitty human.} & high & low \\
			\midrule
			Request & 8 & Hate: \emph{Lmao when you have straight make shit up is when people know you're full of shit.}  &  &\\
			&  & Reply: \emph{What was made up?  Can you articulate your point?} & low & medium\\
			\midrule
			Negation  & 5 & Hate: \emph{He was literally dying, and you motherf**kers laughed at it. } & &\\
			& & Reply: \emph{You seem to imply he was a good person. Doesn't make the jokes any better, but the dude wasn't a role model.} & high & medium\\
			\bottomrule
		\end{tabular}
		\caption{Most common error types made by the best model with predictions made by \emph{hate + reply + blending}.
		}
		\label{t:error}
	\end{table*}

	\noindent
	\textbf{Pretraining with Related Tasks}
	Pretraining could be seen as a two-stage fine-tuning process.
	First, we fine-tune a RoBERTa-base classifier with a related corpus,
	and then with our own corpus.
	We use the following related corpora:
	hate speech~\cite{davidson2017automated},
	counterspeech~\cite{yu-etal-2022-hate},
	stance~\cite{pougue2021debagreement}, and
	sentiment~\cite{rosenthal-etal-2017-semeval}.
	
	\noindent
	\textbf{Blending Additional Data}
	Blending~\cite{shnarch-etal-2018-will} starts combining both a related corpus and our corpus in the fine-tuning process.
	It decreases the portion of instances from the related corpus after each epoch by a fixed ratio~$\alpha$.
	The last $m$ epochs are trained with data only from our own corpus.
	The blending hyperparameters ($\alpha$ and $m$) are tuned like any other hyperparameter (see Appendix~\ref{a:experimentdetails}).
	We use the same corpora as above for pretraining purposes.
	
	\subsection{GPT-4o: Zero- and Few-Shot}
	Having witnessed the success of large language models and prompt engineering~\cite{mishra-etal-2022-reframing},
	we are curious to see whether they outperform supervised approaches using substantially smaller models in our tasks.
	We experiment with GPT-4o using Microsoft Azure API.
	For few-shot prompts, surprisingly, GPT-4o obtains worse results with 4-shot than 1-shot prompting. 
	We refer the reader the Appendices for additional details about the prompts and examples.

	\subsection{Experimental Results}
	Table \ref{t:model-results} shows the results per label and weighted averages. 
	We provide here results pretraining and blending with the most beneficial tasks: stance for pretraining and counterspeech for blending (optimal $\alpha = 1$). 
	Regarding supervised approaches, using only the reply as input offers competitive performance with F1 scores up to 0.47 compared with the random baseline (F1: 0.47 vs. 0.35). 
	Using both the hate comment and the reply as input yields better results (F1: 0.49 vs 0.47).
	Finally, the network that takes both the hate comment and the reply as input and blends the counterspeech corpus or is pretrained with the stance corpus yields the best results (F1: 0.51).
	GPT-4o, however, performs much worse than supervised approaches on our task (F1: 0.42 vs 0.51). 
	
	\subsection{Error Analysis}
	
	While the supervised approaches outperform prompting GPT-4o, 
	forecasting conversation incivility level is a challenging task. 
	We manually analyze 100 randomly sampled errors made by our best model~(\emph{hate+reply+blending}) to reveal the most common error types (Table~\ref{t:error}).
	
	First, 26\% of the errors contain rhetorical questions.
	In the example, the model fails to interpret that the reply is not expecting a reply.
	Instead, the rhetorical question is used to attack the author of the hateful post. 
	Irony or sarcasm are present in 18\% of the errors.
	The reply in the example uses irony (e.g., calling people ``\emph{children}") to humiliate the author of the hateful post, resulting in more incivility in the follow-up conversation (not shown).
	
	More concerningly, some replies contain swear words but they do not elicit additional incivility (13\% of errors).  
	The reply states that the author of the hateful post is a ``shitty human'', and the model considers this would attract more conflicts.
	However, the follow-up conversation has low incivility.
	We also found the model struggles with replies questioning the validity of the content in the hateful post or asking for more evidence~\cite{walton2005fundamentals}.
	Finally, the model errs when there are negations in the replies. 
	This accounts for 5\% of the errors.
	The reply in the example contains multiple negation cues and the model mislabels it.
	
	\section{Conclusion and Discussion}
	In this paper, we work on the task of forecasting incivility of the conversations following replies to hate speech. 
	We have presented new metrics that take into account several dimensions of antisocial and prosocial behaviors in measuring conversation incivility.
	The validation on a human-annotated benchmark demonstrates that
	it is worth accounting for both antisocial and prosocial behaviors,
	although the latter play a smaller role.
	Crucially, our metric 
	aligns with human perceptions more closely than prior work 
	modeling either prosocial or antisocial behaviors,
	or a single dimension of each kind of behavior.
	We find offensive language, explicit hate speech, and abusive language are useful in representing antisocial behaviors.
	Our extensive analyses unpack one of the potential reasons:
	some conversations may include prosocial behaviors, yet people behave differently~%
	(e.g., one person misbehaves regardless of receiving prosocial comments from the others), 
	therefore the overall incivility people perceive is still very high.
	
	Experimental results show that supervised methods outperform prompting LLMs. 
	Specifically, taking into account the hateful posts as well as blending or pretraining with additional corpora yield improvements, 
	yet forecasting future incivility of the conversation following replies to hate speech is still a challenging task. 
	
	\paragraph{Limitations}
	Since we do not conduct randomized controlled experiments, our linguistic analyses should not be interpreted as causal statements.
	Our metrics do not consider the dynamic and multi-layered structure of follow-up conversations.
	Finally, we identify antisocial and prosocial behaviors with classifiers.
	They obtain good results but are not perfect.
	Due to class imbalance in the training data, some classifiers tend to be biased towards the majority class (e.g., not abusive).
	Future research could explore the impact of classifier performance on the conversation incivility metric.

	\bibliography{aaai25}

	\subsection{Paper Checklist}
	
	\begin{enumerate}
		
		\item For most authors...
		\begin{enumerate}
			\item  Would answering this research question advance science without violating social contracts, such as violating privacy norms, perpetuating unfair profiling, exacerbating the socio-economic divide, or implying disrespect to societies or cultures?
			\answerYes{Yes}
			\item Do your main claims in the abstract and introduction accurately reflect the paper's contributions and scope?
			\answerYes{Yes, see Abstract and Introduction}
			\item Do you clarify how the proposed methodological approach is appropriate for the claims made? 
			\answerYes{Yes, see Section 3-6}
			\item Do you clarify what are possible artifacts in the data used, given population-specific distributions?
			\answerYes{Yes, see Section 3-6}
			\item Did you describe the limitations of your work?
			\answerYes{Yes, see the Limitations}
			\item Did you discuss any potential negative societal impacts of your work?
			\answerYes{Yes, see Ethical Statements after the checklist}
			\item Did you discuss any potential misuse of your work?
			\answerYes{Yes, see the Ethical Statements}
			\item Did you describe steps taken to prevent or mitigate potential negative outcomes of the research, such as data and model documentation, data anonymization, responsible release, access control, and the reproducibility of findings?
			\answerYes{Yes, see the Ethical Statements}
			\item Have you read the ethics review guidelines and ensured that your paper conforms to them?
			\answerYes{Yes}
		\end{enumerate}
		
		\item Additionally, if your study involves hypotheses testing...
		\begin{enumerate}
			\item Did you clearly state the assumptions underlying all theoretical results?
			\answerNA{NA}
			\item Have you provided justifications for all theoretical results?
			\answerNA{NA}
			\item Did you discuss competing hypotheses or theories that might challenge or complement your theoretical results?
			\answerNA{NA}
			\item Have you considered alternative mechanisms or explanations that might account for the same outcomes observed in your study?
			\answerNA{NA}
			\item Did you address potential biases or limitations in your theoretical framework?
			\answerNA{NA}
			\item Have you related your theoretical results to the existing literature in social science?
			\answerNA{NA}
			\item Did you discuss the implications of your theoretical results for policy, practice, or further research in the social science domain?
			\answerNA{NA}
		\end{enumerate}
		
		\item Additionally, if you are including theoretical proofs...
		\begin{enumerate}
			\item Did you state the full set of assumptions of all theoretical results?
			\answerNA{NA}
			\item Did you include complete proofs of all theoretical results?
			\answerNA{NA}
		\end{enumerate}
		
		\item Additionally, if you ran machine learning experiments...
		\begin{enumerate}
			\item Did you include the code, data, and instructions needed to reproduce the main experimental results (either in the supplemental material or as a URL)?
			\answerYes{Yes, we will release the code and data upon acceptance.}
			\item Did you specify all the training details (e.g., data splits, hyperparameters, how they were chosen)?
			\answerYes{Yes, see Section 6 and Appendices}
			\item Did you report error bars (e.g., with respect to the random seed after running experiments multiple times)?
			\answerNA{NA}
			\item Did you include the total amount of compute and the type of resources used (e.g., type of GPUs, internal cluster, or cloud provider)?
			\answerYes{Yes, see Appendices}
			\item Do you justify how the proposed evaluation is sufficient and appropriate to the claims made? 
			\answerYes{Yes, see Section 3, Section 6 and Appendices}
			\item Do you discuss what is ``the cost`` of misclassification and fault (in)tolerance?
			\answerYes{Yes, see Section 4 and Limitations}
			
		\end{enumerate}
		
		\item Additionally, if you are using existing assets (e.g., code, data, models) or curating/releasing new assets, \textbf{without compromising anonymity}...
		\begin{enumerate}
			\item If your work uses existing assets, did you cite the creators?
			\answerYes{Yes, see Section 3-6}
			\item Did you mention the license of the assets?
			\answerYes{Yes, see the Ethical Statements}
			\item Did you include any new assets in the supplemental material or as a URL?
			\answerNo{No}
			\item Did you discuss whether and how consent was obtained from people whose data you're using/curating?
			\answerYes{Yes, see the Ethical Statements}
			\item Did you discuss whether the data you are using/curating contains personally identifiable information or offensive content?
			\answerYes{Yes, see the Ethical Statements}
			\item If you are curating or releasing new datasets, did you discuss how you intend to make your datasets FAIR (see \citet{fair})?
			\answerYes{Yes, see Section 3 and Section 4}
			\item If you are curating or releasing new datasets, did you create a Datasheet for the Dataset (see \citet{gebru2021datasheets})? 
			\answerYes{Yes, we have and will release it upon acceptance}
		\end{enumerate}
		
		\item Additionally, if you used crowdsourcing or conducted research with human subjects, \textbf{without compromising anonymity}...
		\begin{enumerate}
			\item Did you include the full text of instructions given to participants and screenshots?
			\answerYes{Yes, see Section 4}
			\item Did you describe any potential participant risks, with mentions of Institutional Review Board (IRB) approvals?
			\answerYes{Yes, see the Ethical Statements}
			\item Did you include the estimated hourly wage paid to participants and the total amount spent on participant compensation?
			\answerYes{Yes, see the Ethical Statements}
			\item Did you discuss how data is stored, shared, and deidentified?
			\answerYes{Yes, see the Ethical Statements}
		\end{enumerate}
		
	\end{enumerate}
	
	\paragraph{Ethical Statements}
	Reddit data are public available. 
	We recognized the public nature of this information does not imply users' consent or willingness to share their data~\cite{fiesler2018participant}. 
	We made the following efforts in protecting personal information of the Reddit community:
	First, we obfuscate user names to avoid identification of specific users. 
	Second, we only report incivility scores of the follow-up conversations after replies to hate speech and do not publish these follow-up conversations.
	Third, upon release of our corpus, we will only release the IDs and labels of replies along with the hateful posts.
	Finally, in compliance with Reddit's policy, we would like to make sure that our dataset will be reused for non-commercial research only.\footnote{\url{https://www.reddit.com/wiki/api-terms/}}
	
	The annotators were warned of the potential uncivil content before working on our task. 
	They were also informed to end the annotation whenever feel uncomfortable or frustrated.
	We provide annotators with access to supporting services throughout the task.
	We acknowledge the risk associated with releasing the corpus, yet we believe the benefit of bringing to light what replies could mitigate future incivility outweighs any risks associated with the corpus release.

	\appendix
	\section{Limitations}
	Our work has four limitations.
	First, we identify antisocial and prosocial behaviors by fine-tuning deep neural networks.
	The performance of these classifiers are satisfactory yet imperfect.
	Second, we only take into account language when forecasting future incivility in conversations following replies to hate speech.
	Other factors, for example user features, may also affect the incivility of follow-up conversations. 
	For example, replying in an uncivil way may not necessarily attract additional hatred to follow, instead, some people may choose not to respond and thus there is no follow-up conversations. 
	Third, we assign equal weights to the selected antisocial (and prosocial) dimensions when aggregating them.
	Although our best metric aligns with human perceptions best compared with prior work, 
	it remains unclear if weighing these dimensions differently could yield improvements or not.
	Finally, we only work with Reddit data in English.
	It is possible that our metrics and findings may not generalize to other language or platforms.
	Addressing these limitations is reserved for future work.

	\section{Additional Details to Identify Antisocial and Prosocial Comments}
	\label{a:classifierperformance}
	
	\begin{table*}[t]
		
		\small
		\centering	
		\begin{tabular}{l ccc ccc ccc}
			\toprule
			\multicolumn{1}{c}{} & \multicolumn{3}{c}{Yes} & \multicolumn{3}{c}{No} & \multicolumn{3}{c}{Weighted Average} \\
			\cmidrule(lr){2-4} \cmidrule(lr){5-7} \cmidrule(lr){8-10} 
			& P & R & F1 & P & R & F1 & P & R & F1  \\
			\midrule
			\addlinespace[2pt]
			RoBERTa training with\\ 		
			~~~~$a_1$ (offensive language) & 0.86 & 0.78 & 0.82 & 0.94 & 0.96 & 0.95 & 0.92 & 0.92 & 0.92\\ 
			~~~~$a_2$ (explicit hate speech) & 0.92 & 0.89 & 0.90 & 0.95 & 0.97 & 0.96 & 0.94 & 0.94 & 0.94\\ 
			~~~~$a_3$ (abusive language) & 0.66 & 0.48 & 0.55 & 0.90 & 0.95 & 0.93 & 0.86 & 0.87 & 0.86\\ 
			~~~~$a_4$ (norm violations) & 1.00 & 1.00 & 1.00 & 1.00 & 1.00 & 1.00 & 1.00 & 1.00 & 1.00 \\ 
			\addlinespace
			~~~~$p_1$ (empathy) & 0.77 & 0.87 & 0.81 & 0.96 & 0.93 & 0.95 & 0.92 & 0.92  & 0.92\\ 
			~~~~$p_2$ (norms) & 0.57 & 0.32 & 0.41 & 0.87 & 0.95 & 0.91 & 0.82 & 0.84 & 0.83\\ 
			~~~~$p_3$ (positiveness) & 0.96 & 0.97 & 0.96 &0.96  &0.96  &0.96  &0.96  & 0.96 & 0.96\\ 
			~~~~$p_4$ (politeness) & 0.73 & 0.53 & 0.62 & 0.86 & 0.94 & 0.90 & 0.83 & 0.84 & 0.83 \\ 
			\bottomrule
			
		\end{tabular}
		\caption{Detailed results obtained with RoBERTa of each antisocial and prosocial behavior.}
		\label{t:classifiers-results}
	\end{table*}
	
	Referring to Section~\ref{sec:corpus}, we use an off-the-shelf RoBERTa-base model (125M parameters) from Hugging
	Face~\cite{wolf-etal-2020-transformers}. 
	We show the performance of each antisocial and prosocial classifiers in Table~\ref{t:classifiers-results}.
	The label is \emph{Yes} when each antisocial or prosocial behaviors is considered as observed by the classifier and \emph{No} when it is not observed by the classifier.
	Note a post is identified as \emph{Yes} in $a_4$ (norm violations) when the content is ``\emph{deleted}'' or ``\emph{removed}'', and thus we did not train a classifier. 
	
	\section{Inter-rater Agreement for Each Group}
	We show the Cohen's $\kappa$ of each group as follows:
	\begin{compactitem}
		\item long vs. long: 0.90
		\item long vs. medium: 0.89
		\item long vs. short: 0.75
		\item short vs. short: 0.74
		\item short vs. medium: 0.79
		\item medium vs. medium: 0.85
	\end{compactitem}
	
	Cohen's $\kappa$ is lower when both follow-up conversations in a pair are short (i.e., short vs. short) or when their length varies a lot (i.e., long vs. short).
	
	\section{Additional Details to Forecast Conversation Incivility Level}
	\label{a:experimentdetails}
	
	\begin{table*}
		\small
		\centering
		\begin{tabular}{lccc}
			\toprule
			& Epochs & Batch size & Learning rate \\
			\midrule
			reply & 5 & 16 & 3e-5 \\
			~~+ pretraining & 3 & 16 & 2e-5\\
			~~+ blending & 3 & 16 & 2e-5 \\
			\bottomrule
		\end{tabular}
		\caption{Hyperparameters used to fine-tune RoBERTa individually for each training setting.}
		\label{t:hyper}
	\end{table*}

	\paragraph{Details and Hyperparameters}
	Referring to Section~\ref{sec:experiments}, we use an off-the-shelf RoBERTa-base model from Hugging Face~\cite{wolf-etal-2020-transformers} to train a classifier that predict conversation incivility levels.
	We run the experiments on a single NVIDIA GeForce RTX 4090 GPU.
	It takes approximately 8 minutes to train one epoch.
	Table~\ref{t:hyper} shows the hyperparameters that yield the highest F1 score in predicting conversation incivility levels.

	\paragraph{Additional Results with FLAN-T5}
	To minimize the variations by different models, we run experiments with FLAN-T5-base using the same experimental setting as RoBERTa-base. 
	Table~\ref{t:model-results-t5} shows the results.
	In general, FLAN-T5 achieve very similar results to RoBERTa on our task.
	
	\paragraph{Experimental Details with LLMs}
	We experiment with GPT-4o and call the API from Microsoft Azure.
	Figure~\ref{f:prompt-zero} shows the zero-shot prompts and Figure~\ref{f:prompt-few} shows the few-shot prompts.
	We set the \emph{temperature} to 0.1 and \emph{top\_p} to 0.1.
	
	\begin{table*}[t]
		
		\small
		\centering	
		\begin{tabular}{l ccc ccc ccc ccc}
			\toprule
			\multicolumn{1}{c}{} & \multicolumn{3}{c}{High} & \multicolumn{3}{c}{Medium} & \multicolumn{3}{c}{Low} & \multicolumn{3}{c}{Weighted Average} \\
			\cmidrule(lr){2-4} \cmidrule(lr){5-7} \cmidrule(lr){8-10} \cmidrule(lr){11-13} 
			& P & R & F1 & P & R & F1 & P & R & F1 & P & R & F1 \\
			\midrule
			\addlinespace[2pt]
			Majority Baseline & 0.00 & 0.00 & 0.00 & 0.52 & 1.00 & 0.68 & 0.00 & 0.00 & 0.00 &  0.27 & 0.52 & 0.35 \\ 
			\midrule
			FLAN-T5 training with \\ 		
			~~~reply & 0.48 & 0.34	& 0.40	& 0.57	& 0.85	& 0.69	& 0.35	& 0.09	& 0.14	& 0.50	& 0.54	& 0.48\\		
			~~~~~~+ pretraining & 0.47	& 0.36	& 0.41	& 0.58	& 0.84	& 0.68	& 0.38	& 0.10	& 0.15	& 0.50	& 0.54	& 0.49\\
			~~~~~~+ blending & 0.44 & 0.43	& 0.43	& 0.59	& 0.76	& 0.66	& 0.37	& 0.15	& 0.22	& 0.50	& 0.53	& 0.50\\ 
			\addlinespace
			
			~~~hate + reply &0 .45 & 0.43 & 0.44 & 0.59 & 0.81 & 0.68 & 0.40 & 0.11 & 0.17 & 0.51 & 0.55 & 0.50\\
			~~~~~~+ pretraining & 0.42 & 0.56 &0.48 & 0.62 & 0.70 & 0.66 & 0.39 & 0.16 & 0.23 & 0.52 & 0.53 & 0.51\\
			~~~~~~+ blending & 0.45 & 0.49 & 0.46 & 0.60 & 0.78 & 0.68 & 0.39 & 0.12 & 0.19 & 0.51 & 0.55 & 0.51\\ 
			\bottomrule
			
		\end{tabular}
		\caption{ Detailed results (F1 score) obtained with FLAN-T5 per label. These results complement Table~\ref{t:model-results}.}
		\label{t:model-results-t5}
	\end{table*}
	
	\begin{figure*}
		\begin{framed}
			\small
			\noindent\texttt{You will be provided with a reply to a hateful post. 
				Will it result in Low, Medium, or High incivility in the follow-up conversation?
				Answer Low, Medium, or High only.
				See below all the possible labels and their description.} \\
			
			\noindent\texttt{Label: Low} \\
			\noindent\texttt{Description: low incivility in the follow-up conversation} \\
			
			\noindent\texttt{Label: Medium} \\
			\noindent\texttt{Description: medium incivility in the follow-up conversation} \\
			
			\noindent\texttt{Label: High} \\
			\noindent\texttt{Description: high incivility in the follow-up conversation} \\
			
			\noindent\texttt{Here is the reply that needs to be classified:} \\
			\noindent\texttt{Reply: ``\textbf{<Reply from corpus>}''}\\
			\noindent\texttt{Label:}
		\end{framed}
		\caption{Template to generate zero-shot prompts for GPT-4o.}
		\label{f:prompt-zero}
	\end{figure*}
	
	\begin{figure*}
		\begin{framed}
			\small
			\noindent\texttt{You will be provided with a reply to a hateful post. 
				Will it result in Low, Medium, or High incivility in the follow-up conversation?
				Answer Low, Medium, or High only.
				See below all the possible labels and their description.} \\
			
			\noindent\texttt{Label: Low} \\
			\noindent\texttt{Description: low incivility in the follow-up conversation} \\
			
			\noindent\texttt{Label: Medium} \\
			\noindent\texttt{Description: medium incivility in the follow-up conversation} \\
			
			\noindent\texttt{Label: High} \\
			\noindent\texttt{Description: high incivility in the follow-up conversation} \\
			
			\noindent\texttt{See below a couple of examples.} \\
			
			\noindent\texttt{Reply: Fuck you centrist bullshit.  You may as well be on the right because you are as just as much a fucking idiot.} \\
			\noindent\texttt{Label: High} \\	
			
			\noindent\texttt{Reply: Well apparently the original black panther comic from the 60s wasn't named after the party.} \\
			\noindent\texttt{Label: Medium} \\	
			
			\noindent\texttt{Reply: Yeah, because modern science says it’s a bit more complicated than that— and confusing the difference between sex and gender doesn’t make him right.} \\
			\noindent\texttt{Label: Low } \\	
			
			\noindent\texttt{Here is the reply that needs to be classified:} \\
			\noindent\texttt{Reply: ``\textbf{<Reply from corpus>}''}\\
			\noindent\texttt{Label:}
		\end{framed}
		\caption{Template to generate few-shot prompts for GPT-4o.}
		\label{f:prompt-few}
	\end{figure*}

\end{document}